\documentclass[ 10pt, a4paper, superscriptaddress]{revtex4}
\usepackage{color} 
\usepackage{epsfig}
\usepackage{csquotes}
 \usepackage{footmisc}
\usepackage{amsmath, amssymb}
\usepackage{hyperref}
\usepackage{graphicx}
\usepackage[capposition=below]{floatrow}
 
\begin{document}
 
           \title{\Large Corrected  thermodynamics of   $(2+1)D$ black hole conformally coupled
to a massless scalar}  
         \author{Himanshu Kumar Sudhanshu}
         \email{himanshu4u84@gmail.com}

 \affiliation{P.G. Department of Physics,   Magadh University, Bodhgaya, Bihar  824234, India}   
 \author{Sudhaker Upadhyay\footnote{Corresponding author}\footnote{Visiting Associate   at Inter-University Centre for Astronomy and Astrophysics (IUCAA),
Pune, Maharashtra 411007, India\label{note3}}}
\email{sudhakerupadhyay@gmail.com}
     \affiliation{Department of Physics, K. L. S. College,  Magadh University,  Nawada, Bihar 805110, India}
  \affiliation{School of Physics, Damghan University, P.O. Box 3671641167, \\ Damghan, Iran}
   \author{Dharm Veer Singh\footref{note3}}
\email{veerdsingh@gmail.com}
\affiliation{ Department Physics,
 GLA University, Mathura 281406 Uttar Pradesh}
  \author{Sunil Kumar}
\email{sunilk21feb@gmail.com}
   \affiliation{Department of Chemistry,  Jagjiwan College, Magadh University, Gaya, Bihar 823003, India} 

\begin{abstract}
We study the corrected entropy due to thermal fluctuation and their effect on the thermodynamics of a conformally 
dressed black hole in three dimensions. We find that the thermal fluctuation affects the entropy significantly for small black holes.   Various corrected thermodynamical variables  
are also calculated for this black hole. We observe that thermal fluctuation on the thermodynamics of a small conformal black hole has a significant effect.
We also analyse the stability of 
black hole under the effect of thermal fluctuation.  Due to thermal fluctuation, instability occurs in the system of small conformal black holes. Isothermal compressibility is also studied for this black hole which diverges for the equilibrium state of the system. 
\end{abstract}

\maketitle

\section{Introduction}
In general relativity (GR),   black holes were first conjectured in $4D$ as a remnant of the gravitational collapse of massive structures \cite{pe}. 
The various black hole solution in $4D$ have been studies \cite{an1,an2,an3,an4}. 
The   black holes studied in dimensions lower than  $4D$ have been found worth in  understanding of associated   physical quantities of the black hole  and in  finding the
matching conformal field theory. 
However, one can not have a black hole solution in $(2 + 1)D$ as the Weyl tensor and Ricci tensor vanishes  
in the absence of matter.   Banados, Teitelboim and Zanelli (BTZ) were the first to discover an exact black hole solution of GR  in  $(2+1)D$   \cite{1,2}.  This   $(2 + 1)D$ black hole solution with constant curvature existed by considering a negative cosmological constant. GR in $(2+1)D$ correlates the $4D$
 Kerr black hole this black hole has features such as mass, temperature, and entropy similar to the Kerr black holes.
 
Later to the discovery of the BTZ black hole, several  $(2+1)D$ black hole solutions
have been explored.  For instance, the three-dimensional black holes where the scalar field  \cite{II1,4,  Singh:2011gd,Singh:2014gva,Singh:2014kaa,Singh:2014apw}  and fermion field \cite{Singh:2014cca} are regular everywhere have been studied. Three-dimensional black holes coupled with generalized dilaton gravity are also analyzed \cite{5}.
A $(2 + 1)D$ static black hole solution with nonlinear electrodynamics satisfying the weak energy conditions is also provided \cite{6}.  An $AdS$   black
hole solution in    $(2 + 1)D$    conformally coupled with scalar and Maxwell fields has been given in Ref. \cite{7}. A $(2 + 1)D$   
charged black holes coupled to natural scalar hair are given in Ref. \cite{8,9}.
Recently, a black hole solution is found for $ f(R)$ gravity  in $(2 + 1)D$ coupled to a self-interacting scalar field  \cite{II2}.

Black holes as a thermal system do not follow the second law of thermodynamics without incorporating the concept of entropy. 
In this connection, it is found that the maximum entropy is related to the event horizon area of black holes \cite{01,II3}. This concept is the basis for holographic principle development \cite{03,04}. 
 In fact, on various occasions, it is confirmed that the maximum entropy of the black holes requires correction, which in turn modifies the holographic
principle \cite{05,06}. The origin for such corrections is the quantum concept of
gravity and of course thermal fluctuations around the equilibrium. These become significant when the black hole becomes smaller due to Hawking radiation.
It has been confirmed that the nature of this correction is logarithmic at the first-order \cite{II5}. These days such correction has been studied and found their importance for several black holes, namely,  quasitopological black holes \cite{15},   charged
rotating   black holes \cite{16},   $f(R)$   black hole \cite{18},    charged massive black holes  \cite{17}, regular black  \cite{b1,b2,b3,Singh:2022izz}, black brane \cite{mir}  and  Horava-Lifshitz black holes \cite{19}. Although thermal fluctuations have been studied considerably but not in the case of the    $(2 + 1)D$ black holes conformally coupled with a scalar. This provides us with an opportunity and motivation 
to fill this gap.   

We first consider the conformal black hole solution coupled with a massless scalar field 
and write a black hole solution and calculate equilibrium thermodynamics for the system.  We know that under various circumstances the entropy of a black hole gets correction which is logarithmic in nature. Consideration of thermal fluctuation is one of them. Here, we derive an exact form of corrected entropy due to thermal fluctuations. This, in turn, modifies all important thermodynamical variables.
Interestingly, we find that the thermal fluctuation  causes instabilities to the system 
for a small black hole. We further calculate isothermal  compressibility for this back hole under thermal fluctuation. Interestingly, a finite value for this   isothermal  compressibility occurs only when there exists thermal fluctuation. In thermal equilibrium, the isothermal compressibility diverges.

In chronological order plan of the paper is as follows. 
In section \ref{sec2}, we provide a brief introduction of  $3D$ conformal black holes coupled with massless scalar field and their equilibrium thermodynamics. In section 
\ref{sec3}, we discuss the first-order corrected entropy of the system due to thermal fluctuation. In section \ref{sec4}, we study the modified thermodynamics of the system due to correction in the entropy. The corrected heat capacity and its effect on the stability of this black hole are presented in section \ref{sec5}. Finally, in section \ref{sec6}, we study the conclusion of the paper. 

\section{Overview of   $(2+1)D$ conformal black hole in}\label{sec2}
In this section, we provide an overview of conformal black holes coupled with massless scalar field  \cite{II1}. The theory is described by the following action: 
\begin{equation}
S=   \int d^{3}x\sqrt{-g}\left[R+\frac{2}{l^{2}}-\nabla_{\mu}\phi\nabla^{\mu}\phi-\frac{1}{8}R\phi^{2}\right],
\end{equation}
where $R$ is  the Ricci scalar, $l$ is length scale related to cosmological constant
and $\phi$ is the (massless) conformal scalar field.
Here, coefficient $8\pi G$ is set to the unit.
 From the variational principle, we get the following field equations corresponding to
 metric and scalar field, respectively \cite{II2}
\begin{eqnarray}
R_{\mu\nu}-\frac{1}{2}g_{\mu\nu}R-g_{\mu\nu}l^{-2}&=&\nabla_{\mu}\phi\nabla_{\nu}\phi-\frac{1}{2}g_{\mu\nu}\nabla^{\alpha}\phi\nabla_{\alpha}\phi+\frac{1}{8}(g_{\mu\nu}\square-\triangledown_{\mu}\triangledown_{\nu}+G_{\mu\nu})\phi^{2},\\
  \label{eqn:II1}
 \square\phi-\frac{1}{8}R\phi &=&0.
\end{eqnarray}
 Here, $\square =\nabla^\mu\nabla_\mu$ is the Laplace  operator.

From the properties of equation (\ref{eqn:II1}), it is obvious that the matter part of the Einstein equation  is traceless and thus leads to constant scalar curvature as
\begin{equation}
\label{eqn:II2}
R=-\frac{6}{l^2}.
\end{equation}

In order to have a black hole solution for the theory, we define  the line element   as follows 
\begin{equation}
ds^{2}=-f(r)dt^{2}+\frac{dr^{2}}{f(r)}+r^{2}d\theta^{2},
\end{equation}
where $f(r)$ is a general metric function. The 
metric function can be specified from the expression of scalar curvature  (\ref{eqn:II2})  as
\begin{equation}
f(r)= \frac{r^2}{l^2}-\frac{a_{1}}{r}-a_{2},
\end{equation}
where $a_{1}$ and $a_{2}$ are constant of integration.
The scalar field $\phi(r)$ can be obtained  from the temporal  and radial components 
of the field equations as
\begin{equation}
\phi(r)=\frac{A}{\sqrt{r+B}},
\end{equation}
where $A$ and $B$ are constant.
The relation between constant of integration $A$, $B$, $a_{1}$ and $a_{2}$ are 
obtained as
\begin{equation}
 a_{1}=\frac{3B^{2}}{l^{2}} ,  a_{2}=\frac{2B^{2}}{l^{2}},\  B=\frac{3}{2}
\frac{a_{2}}{a_{1}},\  A=\sqrt{8B};\  B\geq0.
\end{equation}
 
Thus, the black hole solution is obtained as
\begin{equation}
f(r)=\frac{(r+B)^{2} (r-2B)}{rl^{2}},
\end{equation}
and the the scalar filed configuration will be
\begin{equation}
\phi(r)=\sqrt{\frac{8B}{r+B}}.
\end{equation}

The radius of the event horizon $(r_{+})$ can be determined by solving the root of  
the metric function equation  $f(r)|_{r=r_+}=0$. This gives 
$
r_{+}=2B.
$

Once the metric function is known , it is matter of calculation to determine the 
Bekenstein-Hawking temperature \cite{II3}   from the definition
 \begin{equation}
  T_{H}=\left.\frac{f'(r)}{4\pi}\right|_{r=r_{+}}.
  \end{equation}
   For the obtained solution of conformal black hole in  $(2+1)$-dimensions, the 
   Hawking temperature is obtained as
  \begin{equation}
  \label{eqn:TH}
  T_{H}=\frac{1}{2\pi}\left(\frac{r_{+}}{l^{2}}+\frac{B^{3}}{l^{2}r^{2}_{+}}
  \right)=\frac{9r_{+}}{16\pi l^{2}}.
  \end{equation}
 Now, the important quantity to determine is entropy at the event horizon and this can 
 be calculated using Wald's formula \cite{II4} as
  \begin{equation}
  \label{eqn:S0}
  S_{0}=\frac{A}{4}\left(1-\frac{1}{8} \phi(r_{+})^{2}\right)=\frac{\pi r_{+}}{3},
  \end{equation}
  where $A=2\pi r_{+}$ is the area of Event Horizon.
  \section{The corrected entropy}\label{sec3}
  In this section, we study the corrected entropy of black hole due to small 
  statistical fluctuation and then their effects on the thermodynamics of the system. 
 In order to compute entropy correction, we first define general partition function 
 for the thermal system of black hole as
\begin{equation}
Z(\beta)=\int_{0}^{\infty}\rho(E)e^{-\beta E}dE,\label{par}
\end{equation}
where $\beta= {T_{H}^{-1}}$. The Boltzmann constant is set  unit here.
Now, for this partition function,  the density of state $\rho(E)$ can be derives using 
the inverse Laplace transformation of (\ref{par}) as
\begin{equation}
\label{eqn:II3}
\rho(E) =\frac{1}{2\pi i}\int_{\beta_{0}-i\infty}^{\beta_{0}+i\infty}e^{\mathcal{S}
(\beta)}
d\beta,
\end{equation}
where $\mathcal{S}(\beta)=\ln Z(\beta)+\beta E$ describes the exact entropy of the black hole in 
the form of temperature. At the saddle point, the above complex integral (\ref{eqn:II3}) can be 
calculated by the steepest descent method if the size of black hole are small enough, 
such that $\left(\frac{\partial \mathcal{S}(\beta)}{\partial \beta}\right)_{\beta_{0}}=0$ and $
\frac{\partial^{2}\mathcal{S}}{\partial \beta^{2}}>0$. 

Expanding $\mathcal{S}(\beta)$around the 
equilibrium $\beta=\beta_{0}$, we get
\begin{equation}
\label{eqn:II4}
\mathcal{S}(\beta)=S_{0}+\frac{1}{2}(\beta - \beta_{0})^{2}\left(\frac{\partial^{2} \mathcal{S}}{\partial \beta^{2}}\right)_{\beta_{0}}+.....,  
\end{equation}
where $S_{0}$ denotes to   equilibrium value of   the black hole entropy  (\ref{eqn:S0}). By putting the value of equation (\ref{eqn:II4}) in equation (\ref{eqn:II3}), we get

\begin{equation}
\rho(E)= \frac{e^{S_{0}}}{2 \pi i} \int_{\beta_{0}-i\infty}^{\beta_{0}+i\infty} e^{\frac{1}{2}(\beta - \beta_{0})^{2}(\frac{\partial^{2} S}{\partial \beta^{2}})} d\beta.
\end{equation}
On simplification of the above integral equation, we get the following  density of states as
\begin{equation}
\rho(E) = \frac{e^{S_{0}}}{\sqrt{2 \pi (\frac{\partial^{2}\mathcal{S}}{\partial \beta^{2}})}}
\end{equation}
  The corrected  entropy due to thermal fluctuation can simply be given by taking logarithm of the above density of states  as
\begin{eqnarray}
S&=&\ln \rho (E)\nonumber\\
&=& S_{0}-\frac{1}{2} \ln \left(\frac{\partial^{2}\mathcal{S}}{\partial \beta^{2}}\right).\label{eqn:II5}
\end{eqnarray}
 Here, we have ignored the higher order correction terms.
 For the no work system, it is found that \cite{II5} 
\begin{equation}
\label{eqn:II8}
\frac{\partial^{2}\mathcal{S}}{\partial \beta^{2}}=S_0T_{H}^{2}.
\end{equation}
This leads to corrected entropy (\ref{eqn:II5}) as
\begin{eqnarray}
S&=& S_{0}-\frac{1}{2} \ln  \left(S_{0}T_{H}^{2}\right). \label{zz}
\end{eqnarray}
Thus, the corrected entropy due to thermal correction  has an additional term on the R.H.S. of above expression. To discriminate the contribution of correction term from the equilibrium value, we replace the factor $\frac{1}{2}$ by parameter $\alpha$  in the second term of (\ref{zz}) as
  \begin{equation}
\label{eqn:II9}
S = S_{0}-\alpha \ln \left(S_{0}T_{H}^{2}\right).
\end{equation}
Here, $\alpha$ can  be assigned  only $0$ and $\frac{1}{2}$ values.
For $\alpha=0$, expression (\ref{eqn:II2}) signifies uncorrected entropy 
$S_0$ and, for $\alpha=\frac{1}{2}$,  expression (\ref{eqn:II2}) describes
with corrected entropy (\ref{zz}).
Therefore,  $\alpha$ is called as correction parameter. 

Now, for  the conformal  black hole, thermal fluctuations of entropy around the equilibrium can be estimated. Inserting Hawking temperature $(T_{H})$ and the entropy $(S_{0})$ from equation (\ref{eqn:TH}) and (\ref{eqn:S0}), respectively, in equation (\ref{eqn:II9}), we get the expression for the leading-order correction in entropy due to the thermal fluctuations around the equilibrium of the black hole as follows
\begin{equation}
\label{eqn:S1}
S=\frac{\pi r_{+}}{3}-\alpha \ln \left(\frac{27r_{+}^{3}}{256\pi l^{4}}\right)=\frac{\pi r_{+}}{3}-\alpha \left[ 3\ln(3r_{+})-\ln (256 \pi l^{4})\right].
\end{equation}
Here, from the above expression, we can see that the last two terms represent the  correction to the entropy   due to the thermal fluctuations around the equilibrium. In the limit $\alpha=0$, this corresponds to the equilibrium entropy of the given black hole. In Figure 1, the comparative analysis of the corrected  entropy  with the original one is given.

\begin{figure}[hbt]
    \centering  \includegraphics[width=300pt]{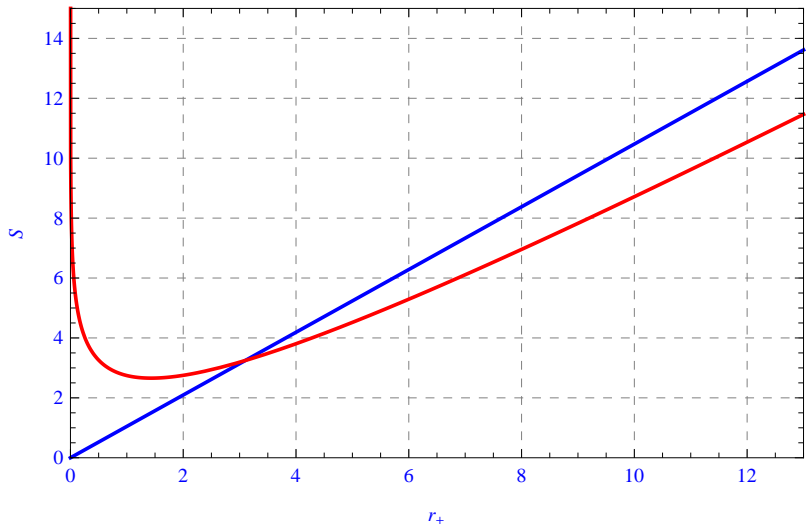} 
  \caption{Variation of entropy $(S)$ with the horizon radius $(r_{+})$ of black hole for $l=1$. Here  $\alpha=0$ (no correction) and $\alpha=  \frac{1}{2}$ (with correction) are denoted by   blue and
    red, respectively.}\label{fg1}
      \end{figure}
      
From the plot (Figure \ref{fg1}), a comparison of corrected and equilibrium entropy for the given black hole  can be drawn. Without correction parameter $(\alpha)$, the equilibrium entropy of the system  is a linearly increasing function as shown by the blue line. For the correction parameter, $\alpha=\frac{1}{2}$, the corrected entropy of the system   always has a positive value which is significant to the system.  The corrected entropy is a decreasing function below the critical horizon radius but above this point, it is always an increasing function. Hence, for the larger black hole, the corrected entropy of the system does not have as significant as expected. But for the smaller black hole system, the correction term due to the small thermal fluctuations has significant effects on the entropy of the system.

\section{A modified thermodynamics} \label{sec4}

\begin{figure}[hbt]
   \centering  \includegraphics[width=300pt]{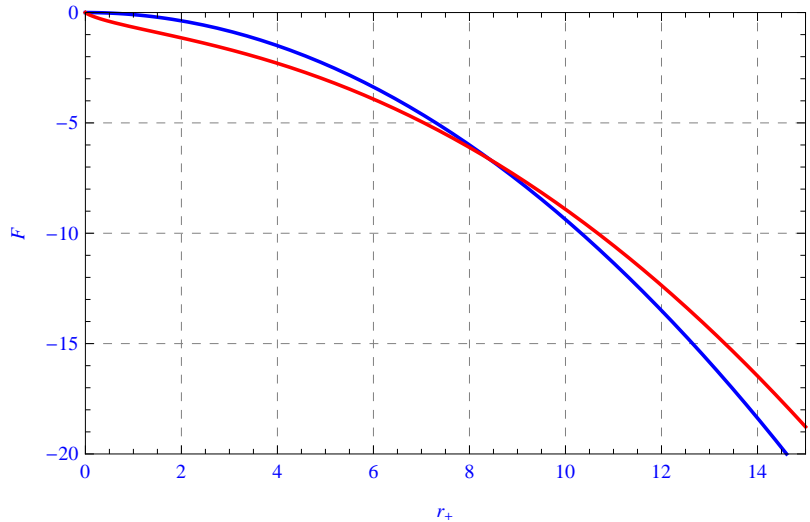} 
  \caption{Variation of Helmholtz free energy $(F)$ with the Horizon radius $(r_{+})$ 
  of black hole for $l=1$.  Here  $\alpha=0$ (no correction) and $\alpha=\frac{1}{2}$ (with correction) are denoted by 
  blue and  red, respectively.} \label{fig2}
   \end{figure}

In this section, we calculate various thermodynamical variables for the black hole system  for the small thermal fluctuations at the equilibrium. Firstly, we would like to calculate the Helmholtz free energy $(F)$. Once we have an expression for the temperature and the entropy, this can be done with the help of the following thermodynamic relation:
\begin{equation}
\label{eqn:F1} 
F=-\int SdT_{H},
\end{equation}
 By plugging the corresponding values of $S$ and $dT_{H}$ from equation (\ref{eqn:TH}) 
 and (\ref{eqn:S1}) in (\ref{eqn:F1}), the exact value for the Helmholtz free energy  
 for the black hole system is calculated by
\begin{equation}
\label{eqn:F2}
F=-\frac{9}{16 \pi l^{2}}\left[\frac{\pi r_{+}^{2}}{6}+\alpha \big\{3 r_{+}-3 r_{+} 
\ln(3r_{+})+ r_{+} \ln(256 \pi l^{4})\big \}\right].
\end{equation}
Now, the behaviour of Helmholtz free energy is depicted in Fig. \ref{fig2}.
 
Here, from this plot, we can do a comparative analysis of    Helmholtz free energy with 
and without thermal correction. The Helmholtz free energy is a decreasing 
function of  horizon radius of the conformal black hole system.
 For the particular (critical) horizon radius, Helmholtz free energy is not affected 
 by the correction parameter i.e. the thermal fluctuation does not affect the system 
 at the critical horizon radius. For  the larger black hole system (with the horizon 
 radius greater than the critical horizon), the corrected Helmholtz free energy has 
 less negative value than the equilibrium values, but for the smaller black hole 
 the system has a reverse behaviour concerning the equilibrium value.  

\begin{figure}[hbt]
    \centering  \includegraphics[width=300pt]{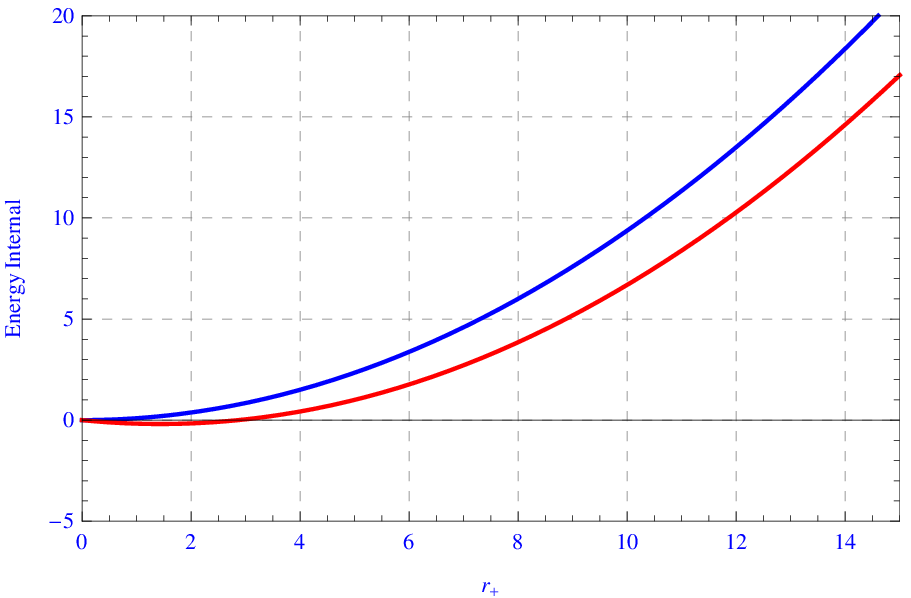} 
    \caption{Variation of internal energy $(E)$ with the hHorizon radius $(r_{+})$ of black hole for $l=1$.  Here  $\alpha=0$ (no correction) and $\alpha=  \frac{1}{2}$ (with correction) are denoted by   blue and  red, respectively.}\label{fig3}
    \end{figure}

The first-law of thermodynamics  corresponding to the corrected entropy $S$ for the 
concerning black hole is given by
\begin{equation} 
dE=T_{H}dS,
 \end{equation}
and this eventually leads to the expression of internal energy (total mass) of the 
system   
\begin{equation}
E=\int T_{H}dS.
\end{equation}
Putting the values of   $T_{H}$ and $dS$ from the equation (\ref{eqn:TH}) and 
(\ref{eqn:S1}) respectively, the  internal energy of the system can be calculated as
\begin{equation}
\label{eqn:E1}
E=\frac{9}{16 \pi l^{2}}\left(\frac{\pi r_{+}^{2}}{6}-3 \alpha r_{+}\right).
\end{equation}
  For the comparative study of the corrected internal energy of the black hole with equilibrium value,  internal energy versus horizon radius has been plotted in figure \ref{fig3}. Here, we find that the internal energy of the black hole increases monotonically with the increase of the horizon radius. Due to the thermal fluctuations, the internal energy decreases  with  the horizon radius of the black hole system.

From the area-entropy theorem of the black hole, the entropy is proportional to the area of the event horizon. Hence,  the volume of the black hole is calculated by
\begin{equation}
\label{eqn:V1}
V=4\int S_{0}dr_{+}=\frac{2}{3}\pi r_{+}^{2}.
\end{equation}

Once we have the expression for the corrected  Helmholtz free energy (\ref{eqn:F2}),
the corrected pressure for the black hole system can be obtained from the expression
\begin{equation}
\label{eqn:P1}
P=-\frac{dF}{dV}=-\left(\frac{dF}{dr_{+}}\right)\left(\frac{dr_{+}}{dV}\right).
\end{equation}
This leads to corrected pressure due to the thermal fluctuations for the black hole system   as
\begin{equation}
\label{eqn:P2}
P=\frac{27}{64 \pi^{2} l^{2}r_{+}}\left[\frac{\pi r_{+}}{3}- \alpha \big\{3 \ln(3r_{+})-  \ln(256 \pi l^{4})\big\}\right].
\end{equation}

\begin{figure}[hbt]
    \centering  \includegraphics[width=300pt]{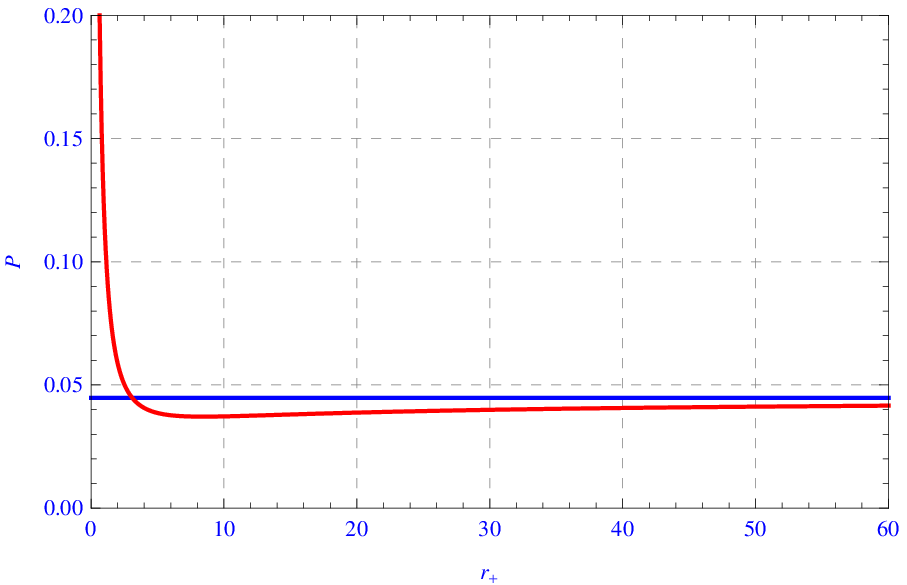} 
    \caption{Variation of pressure $(P)$ with the horizon radius $(r_{+})$ of black hole for $l=1$.  Here  $\alpha=0$ 
    (no correcction) and $\alpha=  \frac{1}{2}$ (with correction)  are denoted by   blue and  red, respectively.}\label{fig4}
    \end{figure}

For the comparative analysis of the corrected pressure due to the thermal fluctuation with the equilibrium pressure of the black hole system, the graph for the corrected pressure  (\ref{eqn:P2})   has been plotted as shown in figure \ref{fig4}. From the plot, we find that for the larger black hole there exists a critical horizon radius beyond which the small thermal fluctuation doesn't have a significant effect on the pressure of the black hole system. Beyond this critical point, corrected pressure due to thermal fluctuations coincides with the equilibrium pressure of the system and becomes saturated. For the larger black hole, small thermal fluctuations don't have a significant effect on pressure, but for the smaller black hole (i.e. smaller than the critical horizon radius) the corrected pressure increases significantly to a very large value due to thermal fluctuations. Hence, for the smaller black hole, thermal fluctuations play a  significant role in the pressure and so can't be ignored. As the correction parameter tends to zero, the equilibrium pressure for the black hole system can be achieved.

The enthalpy $(H)$  is an important parameter for a thermodynamic system, which measures the total heat content of the system.  This can be calculated by the formula 
\begin{equation}
\label{eqn:H1}
H=E+PV.
\end{equation}
 
Putting the value of internal energy $(E)$, pressure $(P)$ and volume $(V)$ from equation (\ref{eqn:E1}), (\ref{eqn:P2}) and (\ref{eqn:V1}) respectively to the above equation (\ref{eqn:H1}), we get the corrected enthalpy as:
which is given by
\begin{equation}
\label{eqn:H2}
H= \frac{3r_{+}^{2}}{16 l^{2}}-\alpha \left(\frac{27 r_{+}}{32 \pi l^{2}}\right)\left[2+ \ln (3r_{+}) - \frac{1}{3} \ln(256 \pi l^{4})\right]
\end{equation}

  \begin{figure}[hbt]
    \centering  \includegraphics[width=300pt]{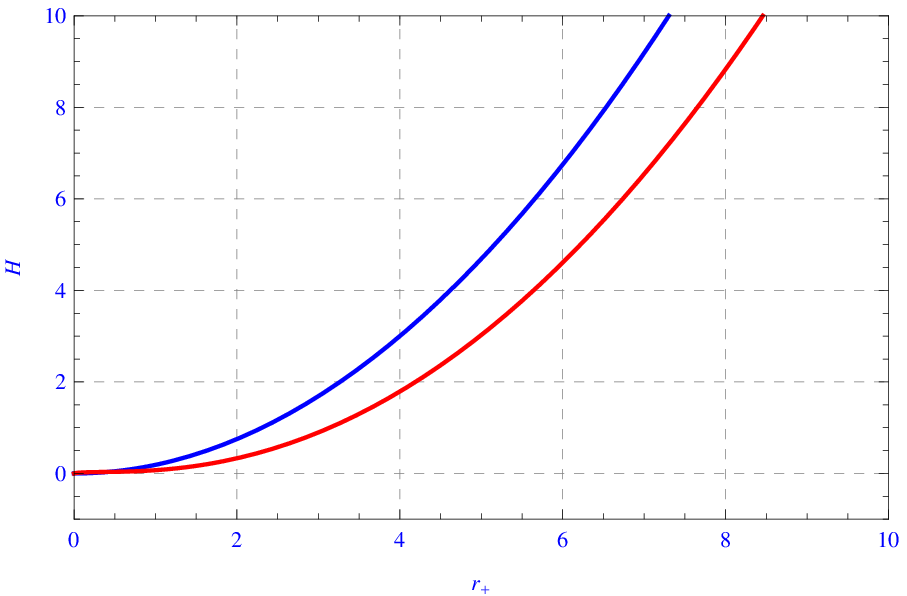} 
    \caption{Variation of enthalphy $(H)$ with the horizon radius $(r_{+})$ of black hole for $l=1$.  Here  $\alpha=0$ (no correction) and $\alpha=  \frac{1}{2}$ (with correction) are denoted by   blue and  red, respectively.}\label{fig5}
    \end{figure}

From the figure \ref{fig5}, the expression of the corrected enthalpy (\ref{eqn:H2}) due to the thermal fluctuations has been plotted with the variation of the horizon radius. From the graph, it can be seen that enthalpy increases with the increase of the horizon radius. Due to the thermal fluctuations, the correction parameter decreases the value of corrected enthalpy in comparison to its equilibrium value.

Another important thermodynamic parameter, Gibbs free energy $(G)$, which tells the maximum work that can be extracted from the thermodynamic system, can be calculated by  
\begin{equation}
\label{eqn:G1}
G=F+PV.
\end{equation}

Substituting the value of Helmholtz free energy $(F)$, pressure $(P)$ and volume $(V)$ from equation (\ref{eqn:F2}), (\ref{eqn:P2}) and (\ref{eqn:V1}), respectively, to the above equation (\ref{eqn:G1}), we get the corrected Gibbs free energy as
\begin{equation}
\label{eqn:G2}
G = -\alpha \left(\frac{9 r_{+}}{32 \pi l^{2}}\right) \left[6-3 \ln (3r_{+}) + \ln(256 \pi l^{4}) \right].
\end{equation}

  \begin{figure}[hbt]
    \centering  \includegraphics[width=300pt]{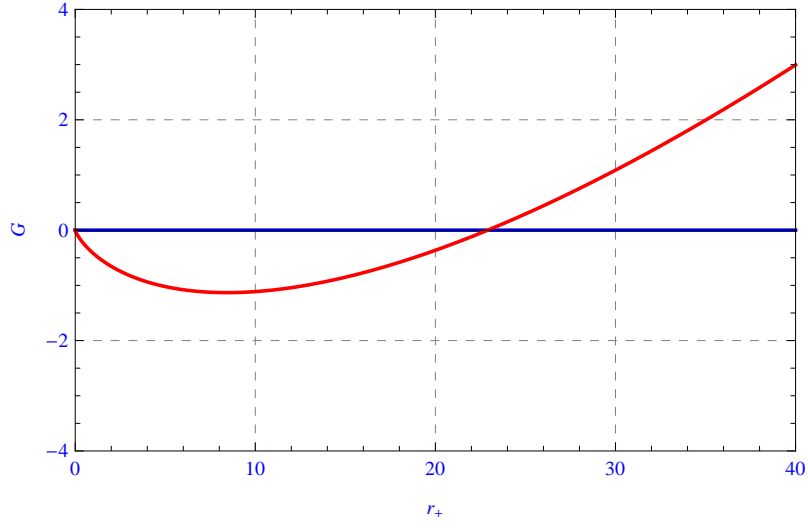} 
    \caption{Variation of Gibbs free energy $(G)$ with the horizon radius $(r_{+})$ of black hole for $l=1$. Here  $\alpha=0$ (no correction) and $\alpha=  \frac{1}{2}$ (with correction) are denoted by   blue and  red, respectively.}\label{fig6}
    \end{figure}
    From the plotted figure \ref{fig6}, wee see that the  Gibbs free energy (\ref{eqn:G2}) vanishes for the black hole in  thermal equilibrium. For the black hole system with thermal fluctuations about the equilibrium, Gibbs free energy has a non-zero value. There exists a critical horizon radius for which the Gibbs free energy changes its sign.  

 \section{Stability of black hole}\label{sec5}
In this section, we study the stability and phase transition of a black hole.   For this 
purpose, we need to calculate the specific heat  of the black hole system.  
Against the phase transition, the stable thermodynamic system has positive specific 
heat otherwise has a negative value.   
 
The specific heat $(C)$ of the thermodynamic system can be calculated by the standard relation as
\begin{equation}
\label{eqn:C1}
C=\frac{dE}{dT_{H}}=\left(\frac{dE}{dr_{+}}\right)\left(\frac{dr_{+}}{dT_{H}}\right).
\end{equation}
Putting the corresponding values from equations (\ref{eqn:TH}) and (\ref{eqn:E1}) to the above equation (\ref{eqn:C1}), we get the corrected specific heat for the black hole under thermal fluctuations at equilibrium as follows 
\begin{equation}
\label{eqn:C2}
C=\frac{\pi r_{+}}{3}-3 \alpha.
\end{equation}

\begin{figure}[hbt]
    \centering  \includegraphics[width=300pt]{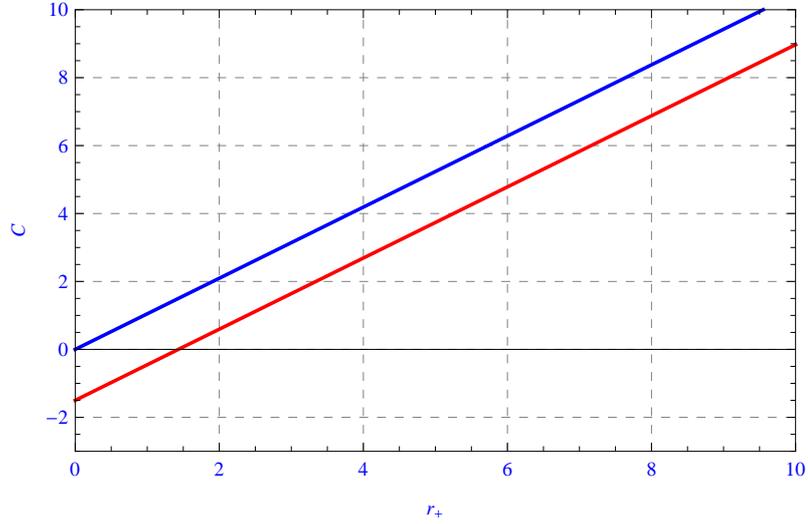} 
    \caption{Variation of specific heat $(C)$ with the horizon radius $(r_{+})$ of black hole for $l=1$. Here  $\alpha=0$ (no correction) and $\alpha=  \frac{1}{2}$ (with correction) are denoted by   blue and  red, respectively.}\label{fig7}
    \end{figure}

In figure \ref{fig7},  the expression of the corrected specific heat (\ref{eqn:C2}) concerning the horizon radius of the black hole has been plotted for the value of the correction parameter. It is clear from the graph that the corrected  heat capacity   has always increasing function with the black hole horizon radius. Specific heat for the equilibrium has always a positive value and hence it is a stable system. But for the smaller black hole, the corrected specific heat due to thermal fluctuations has a negative value, hence it is in an unstable state. Thermal fluctuations at equilibrium don't affect the stability of the larger black hole.

Now, for the black hole system, the effects of the thermal fluctuations at equilibrium on the  isothermal compressibility $(K)$, which measures the change in volume of a black hole concerning the change in the pressure, can be calculated by the formula:
\begin{equation}
\label{eqn:K1}
K=-\frac{1}{V}\left(\frac{dV}{dP}\right)=-\frac{1}{V}\left(\frac{dV}{dr_{+}}\right)\left(\frac{dr_{+}}{dP}\right).
\end{equation}

Putting the corresponding values from equations (\ref{eqn:V1}) and (\ref{eqn:P2}) to the above equation (\ref{eqn:K1}), we get the expression for the corrected isothermal compressibility for the thermal fluctuations at equilibrium as follows: 
\begin{equation}
K= \frac{128 \pi^{2}l^{2}r_{+}}{27 \alpha [3-3\ln3r_{+} + \ln(256 \pi l^{4})]}.
\end{equation}

The above expression for the isothermal compressibility $(K)$ diverges when the   correction parameter $(\alpha)$ tends to zero. Hence at the equilibrium, the black hole system is highly compressible. Therefore, the bulk modulus (inverse of $K$) of the black hole system has zero value at the equilibrium. 
\section{Conclusion}\label{sec6}
GR in three dimensions plays an important role in
understanding the fundamentals of classical and quantum gravity.
Here, we have   discussed the thermodynamics of a conformal black hole solution coupled with a massless scalar field.  On the other hand,  by microstate counting in string theory and loop quantum
gravity, it is confined that   entropy gets logarithmic
correction at first order.  The entropy also gets correction due to small statistical correction around thermal equilibrium.   Due to this correction term, the entropy 
becomes asymptotically large when horizon ceases to a point. 

We have studied the corrected  entropy of  conformal black hole under the influence of thermal fluctuation.  Once there is correction in entropy it is obvious that
the associated thermodynamical variables also receive modification.  
We have first calculated corrected Helmholtz free energy due to thermal fluctuation.
  For  the larger black hole system, the corrected Helmholtz free energy has 
 less negative value than the equilibrium values. However, in contrast, 
corrected Helmholtz free energy becomes more negative than the equilibrium value for the small  black holes. Following the first-law of thermodynamics,
the corrected form of total internal energy is also computed. We have found that the 
thermal fluctuation decreases the internal energy of the system. 
Corrected form of pressure is also calculated. Here, we observed that pressure is not constant if we include thermal fluctuations and pressure increases asymptotically 
when size of black hole reduces to a certain size. The corrected form of enthalpy is also derived. The value of enthalpy decreases due to thermal fluctuations.  
The non-vanishing Gibbs free energy only exists  when we consider thermal fluctuations.  

Interestingly, we found  that, in contrast to equilibrium case, the small conformal black holes are unstable.  The isothermal  compressibility for this back hole is also estimated  under thermal fluctuation. A finite value for    isothermal  compressibility occurs only when there exists thermal fluctuation.  For the  equilibrium state, the isothermal compressibility diverges. 

\section*{Data Availability Statement} 
Data sharing not applicable to this article as no datasets were generated or analysed during the current study.


\begin{thebibliography}{00}
\bibitem{pe}R. Penrose, Phys. Rev. Lett. 14 (1965) 57.
\bibitem{an1}J. M. Z. Pretel, A. Banerjee and A. Pradhan, Eur. Phys. J. C 82 (2022) 180.
\bibitem{an2} T. Tangphati, A. Pradhan, A. Banerjee and G. Panotopoulos, Physics of the Dark Universe 33 (2021) 100877.
\bibitem{an3}T. Tangphati, A. Pradhan, A. Errehymy and  A. Banerjee, Phys. Lett. B 819 (2021) 136423.
\bibitem{an4}G. Panotopoulos, A. Pradhan, T. Tangphati and A. Banerjee, Chin. J. Phys. 77 (2022) 2106.


\bibitem{1}M. Banados, C. Teitelboim and J. Zanelli, Phys. Rev. Lett. 69 (1992) 1849.
\bibitem{2}  M. Banados, M. Henneaux, C. Teitelboim and J. Zanelli, Phys. Rev. D 48 (1993) 1506.
\bibitem{II1}C. Martinez and J. Zanelli,  Phys. Rev. D 54, 3830 (1996).
\bibitem{4} M. Henneaux, C. Martinez, R. Troncoso and J. Zanelli, Phys. Rev. D 65, 104007 (2002).
 \bibitem{Singh:2011gd}D.~V.~Singh and S.~Siwach, Class. Quant. Grav.  {30} (2013) 235034.
 \bibitem{Singh:2014gva}
D.~V.~Singh and S.~Siwach, J. Phys. Conf. Ser. {481} (2014)  012014.
\bibitem{Singh:2014kaa}
D.~V.~Singh, Int. J. Mod. Phys. D {24} (2015)  1550001.
\bibitem{Singh:2014apw}
D.~V.~Singh and S.~Sachan,
 Int. J. Mod. Phys. D  {26} (2016)  1750038.
\bibitem{Singh:2014cca}
D.~V.~Singh and S.~Siwach, Adv. High Energy Phys.  {2015},  528762 (2015).

\bibitem{5} P. M. Sa, A. Kleber and  J. P. S. Lemos,	Class. Quant. Grav. 13 (1996) 125.
\bibitem{6}M. Cataldo, N. Cruz, S. del Campo and A. Garcia,  Phys. Lett. B. 484, 154 (2000).
\bibitem{7}M. Cardenas, O. Fuentealba and C. Martinez,   Phys. Rev. D 90, no.12, 124072 (2014).
\bibitem{8} W. Xu and D. C. Zou,   Gen. Rel. Grav. 49,   73 (2017).
\bibitem{9} W. Xu and L. Zhao,  Phys. Rev. D 87,  124008 (2013).


 
 \bibitem{II2} T. Karakasis, E. Papantonopoulos, Zi-Yu Tang and B. Wang,   Phys. Rev. D 105, 044038 (2022).
 
\bibitem{01} S. W. Hawking, Nature 248, 30 (1974).
  \bibitem{II3} S. W. Hawking,   Commun. Math. Phys. 43, 199 (1975).
\bibitem{03} L. Susskind, J. Math. Phys. 36, 6377 (1995).
\bibitem{04} R. Bousso, Rev. Mod. Phys. 74, 825 (2002).
\bibitem{05} D. Bak and S. J. Rey, Class. Quant. Grav. 17, L1 (2000).
\bibitem{06} S. K. Rama, Phys. Lett. B 457, 268 (1999).
 \bibitem{II5}S. Das, P. Majumdar and R. K. Bhaduri,   Class. Quant. Grav. 19, 2355 (2002). 
\bibitem{15}  S. Upadhyay, Phys. Lett. B 775, 130 (2017).
\bibitem{16} S. Upadhyay, Gen. Rel. Grav. 50, 128 (2018).
  
\bibitem{18} S. Upadhyay, S. H. Hendi, S. Panahiyan and B. E. Panah, Prog. Theor. Exp. Phys.
2018, 093E01 (2018).
\bibitem{17} S. Upadhyay, S. Soroushfar and R. Saffari, Mod. Phys. Lett. A 36, 2150212 (2021).
\bibitem{b1}  B. K. Singh, R. P. Singh, Journal of Physics: Conference Series,  1947(1), 012047 (2021).
 \bibitem{b2} B. K. Singh, R. P. Singh, Journal of Physics: Conference Series,  1947(1), 012010 (2021).
 \bibitem{b3} R. P. Singh,  B. K. Singh, L. K. Sharma, Journal of Physics: Conference Series,   1947(1), 012038 (2021).
\bibitem{Singh:2022izz}
R.~P.~Singh, B.~K.~Singh, B. R.~K.~Gupta and S.~Sachan, Canadian Journal of Physics
100, 1 (2022).
\bibitem{mir} B. Pourhassan and M. Faizal, JHEP 10 (2021) 050.
\bibitem{19} B. Pourhassan, S. Upadhyay, H. Saadat and H. Farahani, Nucl. Phys. B 928, 415
(2018).
 
 \bibitem{II4} R. M. Wald,  Phys. Rev. D 48,  R3427  (1993).
 

 
 \bibitem{II6}J. M. Bardeen, B. Carter and S. W. Hawking,  Commun. Math. Phys. 31, 161 (1973).



 
  \end{thebibliography}
\end{document}